\documentstyle[aps,amsfonts,epsf,12pt]{revtex}

\def\upcite#1{\mbox{$\!^{\cite{#1}}$}}

\newcommand{\bra}[1]{\mbox{$\langle{#1}|$}}
\newcommand{\ket}[1]{\mbox{$|{#1}\rangle$}}

\def\I{{\rm i}}

\def\e{{\rm e}}

\begin{document}
\draft
%\flushbottom
\title{
An Universal Quantum Network -- Quantum CPU
\thanks{Supported by the National Natural Science Foundation of China under Grant No. 69773052 and the Fellowship of China Academy of Sciences}}

\author{An Min WANG$^{1,2,3}$}
\address{CCAST(World Laboratory) P.O.Box 8730, Beijing 100080, People's Republic of China$^1$\\
and Laboratory of Quantum Communication and Quantum Computing\\
University of Science and Technology of China$^2$\\
Department of Modern Physics, University of Science and Technology of China\\
P.O. Box 4, Hefei 230027, People's Republic of China$^3$}

%\date{}

\maketitle

\vspace{0.1in}

\begin{abstract}
%{\begin{center}{\bf ABSTRACT}\end{center}}
{\em An universal quantum network which can implement a general quantum computing is proposed. In this sense, it can be called the quantum central processing unit (QCPU). For a given quantum computing, its realization of QCPU is just its quantum network. QCPU is standard and easy-assemble because it only has two kinds of basic elements and two auxiliary elements. QCPU and its realizations are scalable, that is, they can be connected together, and so they can construct the whole quantum network to implement the general quantum algorithm and quantum simulating procedure}. 
\end{abstract}
\medskip

%\section{Introduction}

The combination of information science and quantum mechanics has created a series of amazing results. One of them is just the idea of quantum computer\upcite{Feynman,Deutsch} which can speed up computation greatly and even exponentially than a classical computer can do, for example, factorization of large number\upcite{Shor} and search for unstructured data.\upcite{Grover} However, there still exist some technologic difficulties beyond our present ability so that a practical quantum computer hasn't been made now. But the rapid developments both in theory and experiments seems to indicate that quantum computer can be implemented in future.

The central part of quantum computer is the quantum network. As is well known, an universal quantum network can be constructed by a set of universal elementary gates\upcite{Barenco} in principle. However, it is still extreme important to know how to construct a whole quantum network to implement a general quantum computing task in practical. They are completely different problems {\it to be able to do} from {\it how to do}. At present, although there are several excellent algorithms, but the knowledge of the whole quantum networks to implement them is still not complete. Moreover, the quantum network for simulating Schr\"odinger equation has not been found. This is because that in terms of Barenco {\it et.al}'s method one does not to know how to assemble and scale up the quantum network for the summation and product of simultaneous and successive transformations. All of this is just my main motivations to write this letter. In fact, I think that an universal quantum network can carry out a general quantum computing task just as a quantum central processing unit (QCPU). Of course, in my point of view,  QCPU should be standard, easy-assemble and scalable, even programmable. As long as it is so, one just can truly make, at least in theory, a quantum computer. 

To arrive at the above aims, I first introduce an auxiliary qubit $A$ prepared in $\ket{0}_A$ and define the universal quantum network for a transformation $U$ as 
\begin{equation}
Q(U)=\prod_{m,n=0}^{2^k-1}\exp\{(U_{mn}\ket{m}\bra{n}\otimes I_A)\cdot C_A^\dagger\},\label{UQN}
\end{equation}
where $I_R$ and $I_A$ are identity matrices in the register space and auxiliary qubit space respectively, and $C_A^\dagger=I_R\otimes c_A^\dagger=I_R\otimes \ket{1}{}_A{}_A\bra{0}$. 
If the graphics rules for the factor with form $\exp\{(U_{mn}\ket{m}\bra{n})\cdot C_A^\dagger\}$ are given out, the picture of quantum network $Q(U)$ can be drawn easily. Obviously, $Q(U)$ is universal and the eq.(\ref{UQN}) is an alternative of Reck {\it et.al}'s formula\upcite{Reck} since it keeps the advantages such as the product form, only involving two states (not qubit) and closed relation with the elementary gates {\it et.al}. Moreover, every factor in $Q(U)$ can be written as an exponential form. Even this construction can be applied to an irreversible and/or non-unitary transformation. In particular, because of its direct corresponding to the transformation matrix, I guess it will be easier to program. Of course, this feature also results in that its assembling is standard and easy. The most important feature is that $Q(U)$ has two new properties
\begin{equation}
Q(U_1+U_2+\cdots+U_r)=Q(U_1)Q(U_2)\cdots Q(U_r),
\end{equation}
\begin{equation}
Q(U_1U_2\cdots U_r)= I_R\otimes I_A+ C_A^\dagger\left(\prod_{j=1}^{r} C_AQ(U_j)\right) C_AC_A^\dagger,\label{CQC}
\end{equation} 
where $C_A$ is so-called ``{\it Connector}" defined as  
\begin{equation}
C_A=I_R\otimes c_A=I_R\otimes \ket{0}{}_A{}_A\bra{1},\label{connector}
\end{equation}
which is used to the preparing transformed state so that this prepared state can be used in the successive transformation. 
Furthermore, note that there are the relations $c_A^2=c_A^{\dagger 2}=0; c_A c_A^\dagger+c_A^\dagger c_A=I_A$, 
thus $c_A$ and $c_A^\dagger$ can be thought as the fermionic annihilate and create operator respectively in auxiliary qubit. From this, the auxiliary system might be able to extended to a larger system beyond a qubit if one can implement easily a nilpotent transiting operator. It is worth pointing out that how to implement the above design is still an open question in experiment and engineering.

Note that in eq.(\ref{CQC}) $I_R\otimes I_A$ is added so that the transformation $U_1U_2\cdots U_r$ is reversible. The another way is to use two registers respectively to input state and out state and the latter includes one auxiliary qubit. Thus, the quantum network for product of transformations becomes a form of full multiplication:
\begin{equation}
\bar{Q}(U_1U_2\cdots U_r)=(I_R)_{input}\otimes \left[C_A^\dagger\left(\prod_{j=1}^{r} C_AQ(U_j)\right) C_AC_A^\dagger\right]_{out},
\end{equation}
while the initial state is now prepared as $(\ket{\Psi(t)})_{input}\otimes (\ket{\Psi(t)}\otimes \ket{0}_A)_{out}$. Therefore, it seems to me this new construction of the universal quantum network is able to scale up easily.

In the above construction (\ref{UQN}), there are two kinds of basic elements $R_m$, $T_{mn}$ defined by
\begin{eqnarray}
R_m(U_{mm})&=&\exp\{(U_{mm}\ket{m}\bra{m}\otimes I_A)\cdot C_A^\dagger\},\\
T_{mn}(U_{mn})&=&\exp\{(U_{mn}\ket{m}\bra{n}\otimes I_A)\cdot C_A^\dagger\}\quad  (m\neq n).
\end{eqnarray}
Obviously, $R_m(U_{mm})$ can be called the ``{\it Rotator}" for its action makes  $\ket{m}\otimes\ket{0}_A$ to rotate to $U_{mm}\ket{m}\otimes \ket{1}_A$. 
$T_{mn}$ can be called the ``{\it Transitor}" for its action makes $\ket{n}\otimes\ket{0}_A$ to transit to $U_{mn}\ket{m}\otimes\ket{1}_A$. 

It is worth emphasizing that there is an essential difference between classical gate and quantum gate. It is just that classical gate is always to carry out a determined operation, but quantum gate can carry out a kind of operations. For example, quantum rotation gate, it can rotate the state to any angle and then it needs one parameter $\phi$ or $\e^{\I\phi}$ to determine its operation. Herewith, so are Rotator and Transitor. 

It is easy to verify that $Q(U)$ acting on $k-$qubit register and single auxiliary qubit $\ket{0}_A$ just implements a general quantum computation:
\begin{equation}
Q(U)\ket{\Psi}\otimes \ket{0}_A=\ket{\Psi}\otimes \ket{0}_A+\ket{\Psi}^\prime\otimes \ket{1}_A,
\end{equation}
where the normalization constant is omitted. In terms of the project measurement $D=I_R\otimes\ket{1}{}_A{}_A\bra{1}$ acting in auxiliary qubit, the transformed result will be obtained. So $D$ can be called ``{\it Drawer}". Another project measurement $I_R\otimes\ket{0}{}_A{}_A\bra{0}$ to carry out the inverse operator of QCPU. Obviously, the reversible operation of $Q(U)$ is
\begin{equation}
Q^{-1}(U) 
=\prod_{m,n=0}^{k-1}\exp\{-(U_{mn}\ket{m}\bra{n}\otimes I_A)\cdot C^\dagger\}.
\end{equation}

In my opinion, the basic elements of QCPU are very fundamental. In mathematics, they are the natural basis of the operator in Hilbert space. Of course, they can be  constructed by the elementary quantum gates.\upcite{Shor2} It is easy to see
\begin{eqnarray}
\ket{m}\bra{n}
&=& \frac{1}{2^k}\prod_{\otimes, i=0}^{k-1}[\delta_{\alpha_i 0}\delta_{\beta_i0}(I+Z)+\delta_{\alpha_i 1}\delta_{\beta_i1}(I-Z)\\ \nonumber
& &+\delta_{\alpha_i 0}\delta_{\beta_i1}(X+Y)+\delta_{\alpha_i 1}\delta_{\beta_i0}(X-Y)];\\ 
\ket{m}&=&\prod_{\otimes, i=0}^{k-1}\ket{\alpha_i},\qquad \bra{n}=\prod_{\otimes, i=0}^{k-1}\bra{\beta_i},
\end{eqnarray} 
where $X=\sigma_x; \I Y=\sigma_y; Z=\sigma_z$ and $\sigma_{x,y,z}$ are usual Pauli spin matrix. In fact, $\ket{m}\bra{n}$ can be written as a product of a general exchange transformation and a measurement $\ket{n}\bra{n}$ from left or a measurement $\ket{m}\bra{m}$ from right. To do this, let's 
introduce the generalized exchange gate $E(m,m+1)$ for two neighbor basic states $\ket{m}$ and $\ket{m+1}$ defined by
\begin{equation}
E(m,m+1)=E(m+1,m)=\sum_{j=0;j\neq m,m+1}^{2^k-1}\ket{j}\bra{j}+\ket{m}\bra{m+1}+\ket{m+1}\bra{m}.
\end{equation}
Obviously, it is Hermian and unitary. It is easy to see that it acts $\ket{m}$ or $\ket{m+1}$ leads to their exchange and is invariant for the other basic states. Note that for two qubits, $E(2,3)$ is a CNOT gate and $E(1,2)$ is a swapping gate.\upcite{Loss} The exchange transformation of arbitrary two states $\ket{m}$ and $\ket{n}$ can be constructed by 
\begin{eqnarray}
E(m,n)&=&\prod_{j=n}^{m-1}E(j+1,j) \quad (n<m),\\
&=&\prod_{j=0}^{n-m-1}E(n-j-1,n-j) \quad (n>m),\\
&=& I_R \quad (n=m).
\end{eqnarray}
The order of product is arranged as left multiplication with index $j$ increasing. Obviously, $E(m,n)\ket{n}=\ket{m}$ and $\bra{m}E(m,n)=\bra{n}$. Therefore, $E(m,n)$ can be expressed by a product of a series of successive the generalized exchange gates in which every the generalized exchange gate is unitary and only involve two neighbor states. 

In the above sense, say it has been designed an universal quantum network which can implement a general quantum computing and keep the advantages in Barenco {\it et.al}'s method. Thus, it can be called the quantum central processing unit (QCPU). 

The key point is efficiency of QCPU. It seems that QCPU consisting of $2^k\times 2^k$ basic elements is the same as the classical computer in use of computing resources. In fact, this is a price to reach at universality. Because the basic elements act on a branch (path) of the quantum data flow, $Q(U)$ needs $2^k\times 2^k$ basic elements. However, many quantum computation tasks have some symmetries, or do not need to act on all states or only act on a subspace, thus it is possible to decrease the numbers of quantum elements largely. For example, the realization of QCPU for a diagonal transformation is $
Q(U_d)=\prod_{m=0}^{k-1}\exp\{(U_{m}\ket{m}\bra{m}\otimes I_A)\cdot C^\dagger\}$. 
Moreover, Rotator, Transitor, even their combination with many branches can be introduced. For example, the conditional rotation gate $R_2$ only acting on the second qubit in 3-qubit register is made of two Rotators with $2^3/2$ branches.\upcite{Ekert}   
Generally speaking, the number of the elements of a realization of QCPU for the transformation $U$ is at least equal to the number of the inequality matrix elements in $U$ expect for zero. 

The simplest realization of QCPU is $Q(0)=I_R\otimes I_A$ for zero matrix or $Q(I_R)=I_R\otimes (I_A+\ket{1}{}_A{}_A\bra{0})$ for identity matrix. Another typical case is the computing step $V$ can be written as the direct product $V(1)\otimes V(2)$. Suppose $V(1)$ is $2^{k_1}\times 2^{k_1}$ and $V(2)$ is $2^{k_2}\times 2^{k_2}$ $(k_1+k_2=k)$. Then, if in $V(1)$ we can decrease a parameter or an elements, the result leads that $2^{k_2}\times 2^{k_2}$ parameters or the number of the corresponding elements are decreased. The use of computing resource is then at high efficiency. For example, a conditional rotation only for the $i-$th qubit, the transformation matrix is $R_i(\phi)=I_1\otimes\cdots\otimes I_{i-1}\otimes (\ket{0}\bra{0}+ \e^{\I\phi}\ket{1}{1})\otimes I_{i+1}\otimes\cdots\otimes I_{k}$. $Q(R_i)$ is then a Rotator with one parameter and all of branches. 
So, to simplify $U$ into the direct product of subspaces as possible is a better method to advance the efficiency of the use of computing resources. 
Quantum Fourier transformation is just an example. In my paper\upcite{My3}, I have obtained its quantum network on a $k-$qubit register:
\begin{equation}
Q(F)=\prod_{n=0}^{2^k-1}Q[B(n)HM_{0n}]= \prod_{m=0}^{2^k-1}\prod_{n=0}^{2^k-1}\exp\{[(B(n)H)_{m0}E(m,n)\ket{n}\bra{n}\otimes I_A] \cdot C^\dagger\},
\end{equation}
where $B(n)H=\prod_{\otimes,j=0}^{2^k-1}B_j[2^j\pi n/(2^k-1)]]H_j$.
While $M_{0n}=\ket{0}\bra{n}$. 

It necessary to point out that there is an essential difference in classical computing and quantum computing. In a classical algorithm, one usually does not need to consider how to connect two computing steps because the classical data flow is in general single branch (unless in parallel). But, in a quantum algorithm, one has to think over this problem because the quantum data flow is in general many branches. If with respect to two quantum computing steps, that is two unitary transformations, one designs their quantum networks independently, then the arrangement ways of quantum data of input and output for two quantum networks are different in general. This means that an interface unit is needed. How to design this interface unit just becomes a problem. Here, in my construction of QCPU, Connector as a standard interface unit is given out and one does not worry about this problem again. For example, the quantum network for time evolution operator can be constructed as:
\begin{equation}
\bar{Q}(\e^{\I Ht})= (I_R)_{input}\otimes\left[C_A^\dagger\left(\prod_{i=1}^{[t/\triangle t]} C_A Q(\Omega(\triangle t))\right)C_AC_A^\dagger\right]_{out}, \label{SEQN}
\end{equation}
where the quantum network for the evolution operator $\Omega(\triangle t)$ in a short time step $\triangle t$ then reads 
\begin{equation}
Q(\Omega(\triangle t))=Q(I_R)Q(\I\triangle t H)=Q(I_R)Q(\I\triangle t T)Q(\I\triangle t V).
\end{equation}
By using of the eq.(\ref{SEQN}), Schr\"odinger equation can be simulated in general.      

In summary, I proposed a new construction of the universal quantum network -- quantum CPU, which is consist of two basic elements: Rotator and Transitor, and two auxiliary elements: Drawer and Connector. The elements of QCPU are standard and very easy to assemble and scale up. Because the connector is designed, the whole quantum network can be obtained conveniently for a quantum algorithm including quantum simulating and then it can process the complex computing task with many computing steps. From this QCPU proposed by this letter, it follows a general principle of design of quantum algorithm including quantum simulating procedure. That is, for a quantum computing task or an unitary transformation $U$, all we need to do for design of quantum algorithm and quantum simulating procedure is to seek a its optimized decomposition seriation of transformations including quantum measurements in an appropriate space. Meanwhile, this seriation should be the most suitable and the realization of QCPU of every step (quantum transformation) is the simplest. In order to do this, we need to use the fundamental laws of physics, specially the principles and features of quantum mechanics, for example, coordinate system choice, representation transformation, picture scheme and quantum measurement theory, if we have thought that a quantum computing task is a physical process. Moreover, to simplify the realization of QCPU for all computing steps and find the optimized decomposition, we have to use the symmetry property of every step $U^i$ as possible, such as the direct product decomposition, the transposing invariance $U_{nm}^i=U_{mn}^i$ or the row equality $U_{mn}^i=U_{m0}^i$ for all $n$ or the line equality $U_{mn}^i=U_{0n}^i$ for all $m$ as well as make $U_i$ with zero elements and equal elements as many as possible.
Limited the space, I can not give the concrete example here. But, I will give them and the applications of QCPU in my papers.\upcite{My2,My3} In those papers, I will describe how to simulate Schr\"odinger equation in general and how to implement Deutsch algorithm, quantum Fourier transformation, Shor's algorithm and Grover's algorithm. 
 
\bigskip
  
I would like to thank Artur Ekert for his great help and for his hosting my visit to center of quantum computing in Oxford University. I also thank Markus Grassl for his helpful comments.


\begin{references}
\bibitem{Feynman} R. P. Feynman. {\it Int. J. Theor. Phys.} {\bf 21} (1982)467 
\bibitem{Deutsch} D. Deutsch. {\it Proc. R. Soc. Lond. A} {\bf 400} (1985)97 
\bibitem{Shor}P. W. Shor.  in {\it Proceedings of the 35th Annual Symposium on the Foundations of Computer Science}, IEEE Computer Society Press, Los Alamitos, CA. ed. S. Goldwasser. 1994. 20; {\it SIAM Journal of Computation} {\bf 26} (1997)1484 
\bibitem{Grover} L. K. Grover. in {\it Proceedings, 28th Annual ACM Symposium on the Theory of Computing}. 1996. 122, preprint quant-ph/9605034; {\it Phys. Rev. Lett.} {\bf 79} (1997)325
\bibitem{Barenco} A. Barenco {\it et. al}, ``Elementary gates for quantum computing", quant-ph/9503016
\bibitem{Reck} M Reck, A. Zeilinger, H. J. Bernstein and P. Bertani, {\it Phy. Rev. Lett.} {\bf 73} (1994)58
\bibitem{Nielsen}M. A. Nielsen and I. L. Chuang ``Programmable quantum gate arrays", quant-ph/9703032
\bibitem{Shor2} P. W. Shor. in {\it Proceedings of the 37th Annual Symposium on the Foundations of Computer Science}, IEEE Computer Society Press, Los Alamitos, CA. 1996. 15
\bibitem{Loss} D. Loss and D. P. DiVincenzo, {\it Phys. Rev. A} {\bf 57} 120(1998) 
\bibitem{Ekert} A. Ekert, Phys. Scripta {\bf T76} 218 (1998); R. Cleve, A. Ekert, C.Macchiavello and M. Mosca, {\it Proc. R. Soc. Lond. A} {\bf 454} (1998)339 
\bibitem{My2} An Min Wang, ``Quantum CPU and quantum simulating", preprint quant-ph/9910090
\bibitem{My3} An Min Wang, ``Quantum CPU and quantum algorithm", preprint quant-ph/9910091

\end{references}
\end{document}